\documentstyle[11pt,aaspp4]{article}

\tighten

\newcommand{\gtsim}{\ {\raise-0.5ex\hbox{$\buildrel>\over\sim$}}\ }
\newcommand{\ltsim}{\ {\raise-0.5ex\hbox{$\buildrel<\over\sim$}}\ }

\def\simlt{\lower.5ex\hbox{$\; \buildrel < \over \sim \;$}}
\def\simgt{\lower.5ex\hbox{$\; \buildrel > \over \sim \;$}}

\begin{document}

%alternate title
%\title{The End of Passive Evolution of Elliptical Galaxies
%from Deep Optical and Near-Infrared Imaging\\

\vspace{12pt}

\title{The Absence of Passively Evolving Ellipticals in Deep Optical
and Near-Infrared Surveys}

\vskip 6pt
\author{Stephen E. Zepf$^{1,2}$}

\vskip 6pt
\noindent $^{\rm 1}$ Department of Astronomy, University of California, 
Berkeley, CA 94720-3411

\noindent $^{\rm 2}$ Department of Astronomy, Yale University,  
New Haven, CT 06520-8101

%\begin{abstract}

\vskip 12pt

{\bf The traditional view of elliptical
galaxies has been that they formed in a single, rapid burst of star
formation at high redshift, and have evolved quiescently since
that time. In opposition to this traditional view is evidence
that at least some elliptical galaxies have formed from the 
merger of two disk galaxies. What has not been clear is which 
process is the dominant formation mechanism for the large majority
of elliptical galaxies. This question has significant implications
for cosmological models, as different models make different 
predictions for the formation mechanism and epoch of elliptical 
galaxies$^{1}$.
Here I use deep optical$^{2}$ and near-infrared$^{3-6}$ images 
% - could be williams et al., Hogg, Dickinson, Moustakas, and Cowie. !!!
to show that there are fewer galaxies with very red optical and
near-infrared colors than predicted by models in which typical
elliptical galaxies have completed their star formation by 
$z \gtsim 5$. 
These observations require that elliptical galaxies have significant star 
formation at $z < 5$.
This requirement, combined with constraints on
lower redshift starbursts from the modest ultraviolet luminosities of
galaxies in the Hubble Deep Field$^{7,8}$ and the properties of galaxies 
from $0 < z < 1$ (refs 9-13), suggests that elliptical galaxies form 
at moderate redshift in dusty starbursts and/or through the 
hierarchical merging of smaller objects.}

\vskip 12pt

	Very deep images at optical and near-infrared wavelengths
provide a sensitive test of models of elliptical galaxy evolution.
Passive models in which ellipticals form in a single, rapid burst
at high redshift predict the existence of a population of high
redshift galaxies with extremely red colors. In particular, 
galaxies that have formed the bulk of their stars at high redshifts 
will have extremely red optical and near-infrared colors when observed 
at ($ 1 \ltsim z < \ z_f$), where $z_f$ is the formation redshift.
The red colors are the result of the absence of hot, 
young massive stars combined with the effect of redshift which causes 
the observed near-infrared and optical colors to become redder as they 
probe farther into the restframe blue and ultraviolet. This trend towards 
redder colors at higher redshifts only changes as the redshift of 
formation is approached, when the hot and blue massive stars 
have not yet died out.

        The predictions of these models are shown in Figure 1,
which traces the colors of galaxies formed in bursts of
star formation at various redshifts for representative cosmologies.
This diagram demonstrates that galaxies which 
have completed a large majority of their star formation by $z \simeq 5$ 
reach extremely red colors of V$_{606} - {\rm K} > 7$. Galaxies with 
significant star formation at lower redshifts do not have these colors, 
unless reddened greatly by dust.
Therefore, the number of objects with V$_{606} - {\rm K} > 7$
gives an upper limit to the comoving number density of passively
evolving elliptical galaxies with $z_f \gtsim 5$.

	Only recently have surveys become deep enough in both the optical
and near-infrared bands to test for the presence of these extremely
red galaxies at $z > 1$. This advance is due in large part to the
optical images of the Hubble Deep Field (HDF) $^{2}$ that allow 
detections or useful upper limits on objects with the very faint V 
magnitudes expected of red objects at these redshifts. These optical
images have been followed up by several near-infrared surveys
that can be used to determine the number of galaxies in the HDF 
with very red V$_{606} - $K colors. One of these surveys provides
K magnitudes to very faint limits for about $20\%$ of the HDF$^{3}$.
Near-infrared images of the entire HDF to a somewhat brighter limit
have also been obtained and made publically available $^{4}$, 
and these have been analyzed as part of this work.
Two ground-based surveys in different fields provide
additional constraints on the the number of extremely red
galaxies$^{6,7}$. Although these ground-based surveys are
not as deep, as the HDF surveys, particularly at optical
wavelengths, they are valuable because they are in completely
independent fields, widely separated on the sky.

	The results of these surveys are compared to the predictions
of passively evolving models in Figure 2.  The surface density
of extremely red galaxies objects is far below the expectations
of a model in which all early-type galaxies form in bursts at
high redshift. The predictions for the surface density of extremely
red galaxies are calculated using a constant comoving number density
of early-type galaxies set by studies of the
K-band luminosity function at low redshift$^{\rm 14}$,
the stellar populations models discussed
earlier, and the appropriate cosmology for each plot.
This calculation may underestimate the number density of extremely red
galaxies expected in passively evolving models because the adopted
local luminosity function has a density normalization which
is about a factor of two lower than more global
measures at modest redshifts$^{\rm 15}$.
Alternatively, the calculation may be an overestimate in that
the luminosity function used includes both ellipticals and S0s,
whereas passive models could be said to only apply to ellipticals,
which are about $50\%$ of the total.
	
	The strong deficit of galaxies with extremely red colors,
seen in widely separated fields and at different flux limits,
rules out models in which typical elliptical galaxies are fully
assembled and have formed all of their stars at $z \gtsim 5$.
This constraint cannot be escaped simply by breaking ellipticals 
into many pieces which formed their stars at high redshift.
Near-infrared surveys are now deep enough to detect
the individual pieces, and thus such models dramatically
overproduce red galaxies within the observed range of K magnitudes.
An illustrative example with $z_f=5$, 
a constant total stellar mass in the early-type population, 
and a comoving number density of this population that increases as $(1+z)^3$,
is plotted in Figure 2. Models with more modest density evolution
follow curves between this model and the $z_f=5$ passive model with
constant number density.
{\it Therefore, typical elliptical galaxies must have had
significant star formation at $z < 5$.}

	Single-burst models at lower redshifts are strongly constrained 
by other observations. In particular, the modest ultraviolet luminosities 
observed for galaxies in the HDF$^{7,8}$, as well as the 
failure of searches for strong emission-line objects$^{\rm 16}$
rule out these models, unless the objects are obscured greatly 
by dust$^{\rm 17,18}$. Dusty models produce high luminosities in the 
far-infrared and sub-mm, and are close to current observational
limits at these wavelengths$^{\rm 7}$.
%for both individual sources and integrated background

	Alternatively, the simple passive model at high redshift can 
be altered to incorporate a much longer starburst duration that
extends to lower redshift. In order to make the colors sufficiently
blue, this component must form about five percent or more of the 
stellar mass of ellipticals from $z_f$ through $z=1$. For example, 
if $5\%$ of the stellar mass forms in an additional star formation 
component that is constant from $z_f$ to $z=1$ and then shuts off, 
the resulting V$_{606} - {\rm K}$ color is slightly redder than 5 
over most of this redshift range. If the mass in the extended component 
is only a few percent, then the V$_{606} - {\rm K}$ color is greater 
than 6, and far more red galaxies are expected than observed.
Thus, the extended star formation must represent 
%a significant fraction of 
at least $5\%$ of the stellar mass to accommodate the absence of 
very red galaxies.
This modification of the passive model runs into several problems 
with other observational constraints. Firstly, it increases the
evolution to brighter luminosities at higher redshifts expected 
in passive models, contrary to the results of deep, red-selected 
galaxy redshift surveys which find little or no luminosity evolution$^{9-11}$.
Moreover, an extended star formation history is unable to account 
for the elemental abundance ratios in ellipticals$^{\rm 12}$.

	A more natural answer is that most elliptical galaxies
form at $z < 5$ through merging and associated starbursts,
which are likely to be at least moderately dusty. 
This result is consistent with a large body of other evidence 
which indicates that merging plays a major role in the formation of 
ellipticals, ranging from detailed studies of nearby galaxy mergers 
to the discovery of bimodality in the globular cluster systems of 
elliptical galaxies$^{\rm 13}$. 
The formation of elliptical galaxies in this way is consistent with 
the predictions of hierarchical clustering models of galaxy formation
$^{19-21}$.

{\noindent ACKNOWLEDGEMENTS.
	I acknowledge stimulating conversations with colleagues at Berkeley, 
Cambridge, and Durham. I thank L.\ Moustakas for assisting with the use
of image analysis software, R.\ Bouwens for providing code to manipulate 
stellar populations models, and D. Hogg for providing data in tabular form.
This work has been supported by NASA grants.}

{\noindent CORRESPONDENCE should be addresses to S.E.Z. 
(e-mail zepf@astro.yale.edu).

\newpage

\centerline{\bf Figure Captions}

\vskip 12pt
\noindent{Figure 1. A plot of the observed V$_{606} -$ K color as a function
of redshift for a stellar population formed in a single short burst
($\tau = 0.01$ Gyr). Galaxies with higher formation redshifts ($z_f \simgt 5$)
have extremely red colors that are not reached by galaxies with lower
formation redshifts ($z_f \simlt 3$). The arrow marks the color cut
used in Figure 2. Both the closed model plotted in the top panel and 
the open model in the bottom panel exhibit the same effect. An approximate 
translation from the HST V$_{606}$ bandpass to the standard  V bandpass 
is given on the right side of the plots. These figures 
are based on the Bruzual \& Charlot models (manuscript in preparation)
with solar metallicity and a Miller-Scalo stellar initial mass function (IMF).
The results are not sensitive to the IMF within
the usual range of parameters (e.g.\ Salpeter vs.\ Miller-Scalo), and
a short burst duration is used only to isolate the effect of the formation 
redshift. For simplicity, absorption by intervening HI has not been
accounted for$^{22}$, which will only make the colors redder at $z \gtsim 3$.}

\vskip 4pt

\noindent{Figure 2. The surface density of extremely red galaxies
(V$_{606} - {\rm K} >7$) at different limiting K magnitudes
compared to the predictions of passive models of elliptical
galaxy formation. The data points are inconsistent with $z_f \gtsim 5$ 
models, including the light dotted line representing a pure density 
evolution model with $z_f = 5$. The two circles represent surveys 
in HDF. The deeper of these is the Hogg et al.\ survey with 
the Keck telescope of roughly one arcmin$^2$ to a $50\%$ completeness
limit of $K \approx 23.5$, in which no objects with colors red 
as V$_{606} - {\rm K} > 7.0$ are found, 
and only one is even within one magnitude of this color$^{3}$.
The second HDF data point is based on the publically
available Kitt Peak imaging survey$^{4}$ that covers the full
five arcmin$^2$ of the HDF, to a $50\%$ completeness limit of 
K$ \approx 22.25$.
Using an updated version of the SExtractor image analysis program$^{\rm 23}$,
we find that all sources with $K < 22$ are detected 
in the HDF optical images and have significantly bluer colors than 
the passive model predictions. At the fainter limit
of K$ < 22.5$, there is one good candidate and two marginal candidates
for objects with V$_{606} - {\rm K} > 7.0$. The point plotted accounts
for all three of these candidates, and is therefore shown as an 
upper limit. The square and triangle are from
ground-based surveys in other fields by Moustakas et al.\ $^{5}$,
and Cowie et al.\ $^{6}$ respectively. 
The former covers approximately two arcmin$^2$ to a $50\%$ completeness
limit of K$ \approx 22.5$. Within these limits, there are three
galaxies that are undetected in V, with approximate lower limits to 
their (V$-$K) color of 7.5, 6.1, and 5.3 (ref.\ 5). These three
represent an upper limit to the number of extremely red objects
in this field.
The latter survey covers about six arcmin$^2$ to a 
limiting K magnitude of 20.9. Because the bluer bandpass is not very
deep in these data, I adopt (I$-$K)$> 4.5$ as the best estimate of the 
equivalent V$_{606} - {\rm K} > 7$ cut used above. There are approximately
nine galaxies with colors this red. This number is more properly considered 
an upper limit, as there are several objects
with red (I$-$K) but blue optical colors$^{5,6}$.}

\newpage

\begin{figure}\plotone{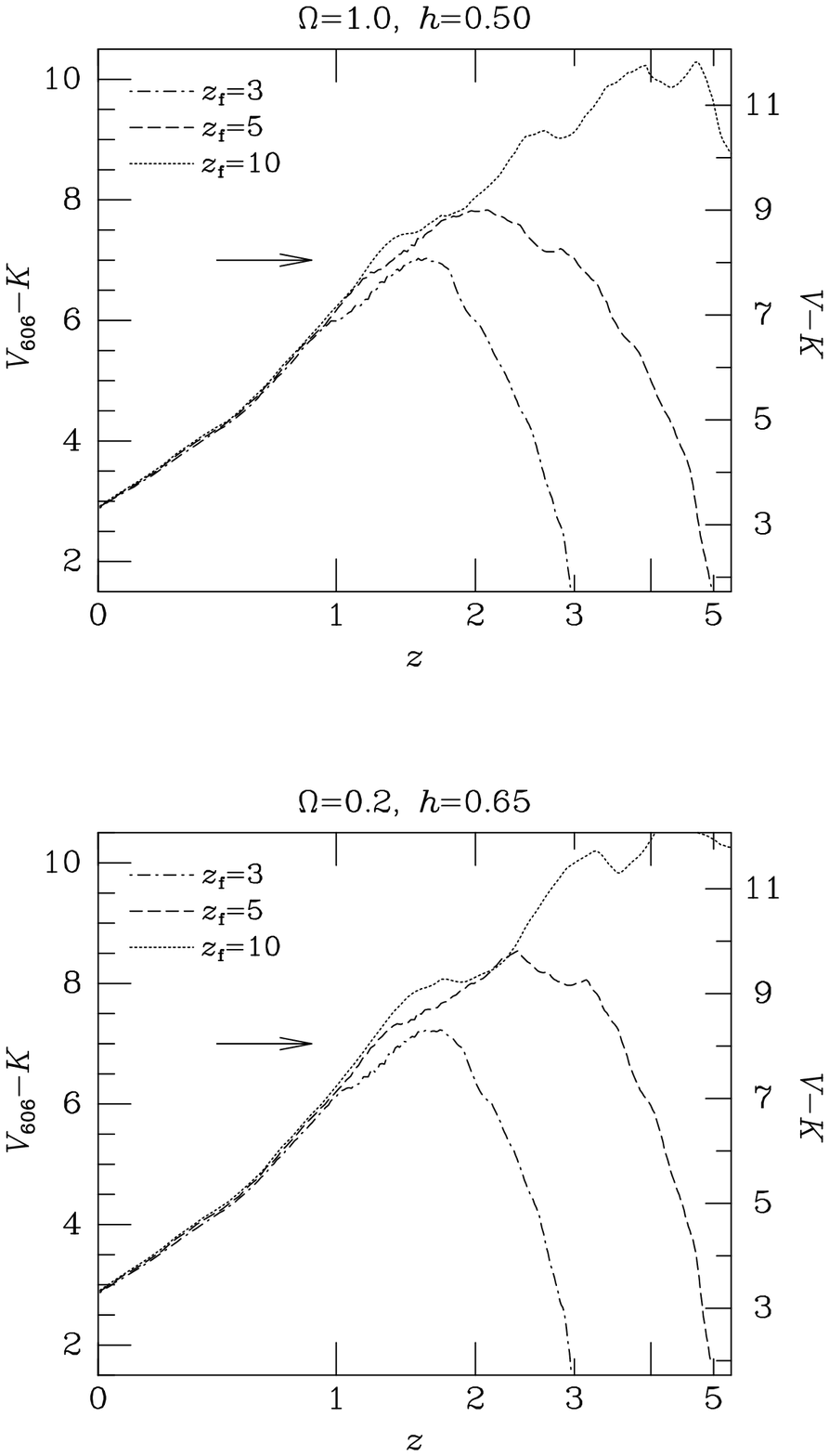}
\end{figure}

\begin{figure}
\plotone{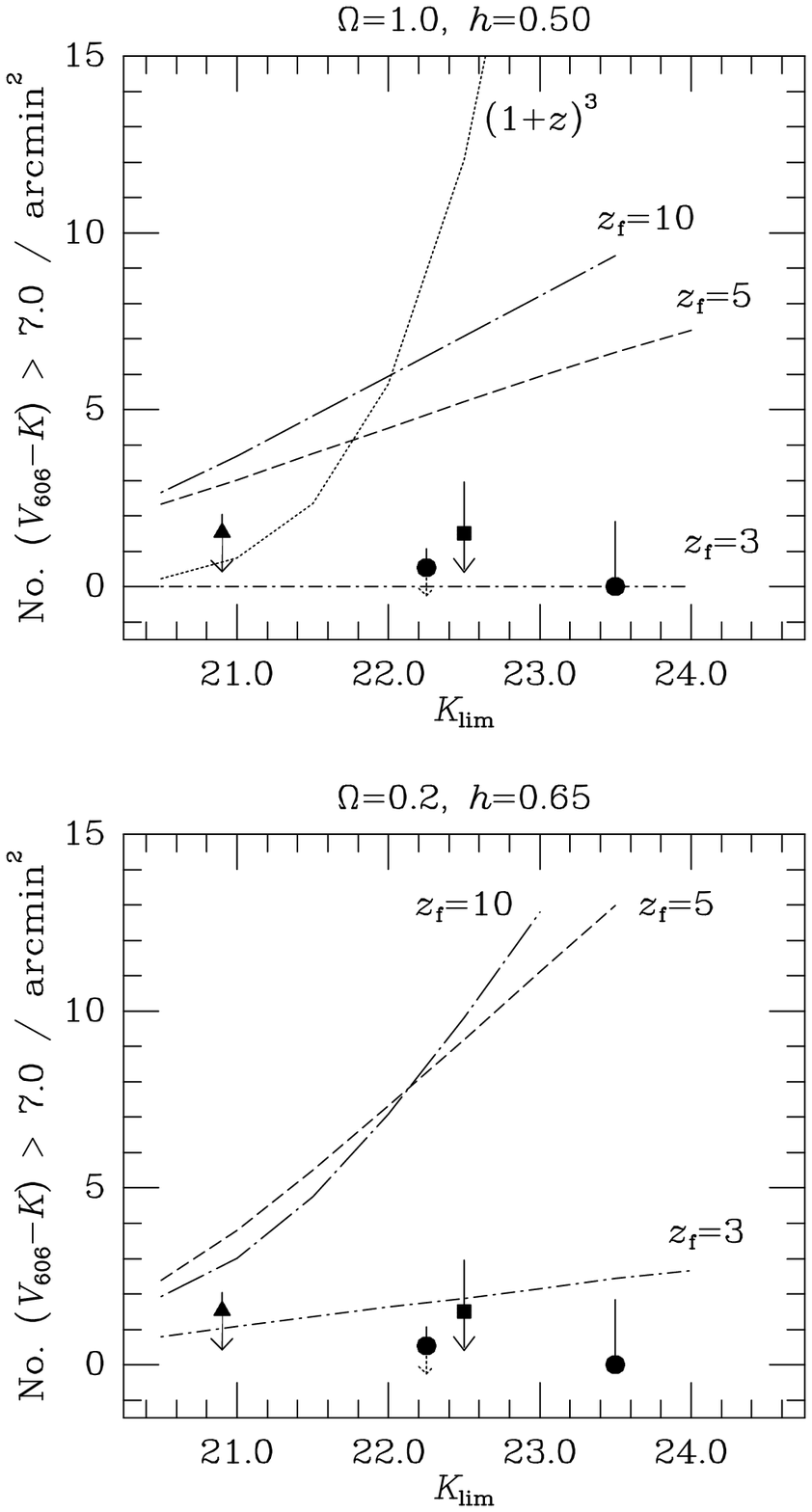}
\end{figure}

\end{document}